\documentstyle[12pt]{article}

\def\simle{{}^<_{\sim}}
\def\simge{{}^>_{\sim}}
\begin{document}
\baselineskip24pt
\pagestyle{plain}

\begin{center}
\vspace{0.2in}
{\Large \bf Primordial Black Hole Formation}\\
\bigskip
{\Large \bf during the QCD Epoch}\\
\vspace{.3in}

Karsten Jedamzik

Physics Research Program

Institute for Geophysics and Planetary Physics

University of California

Lawrence Livermore National Laboratory

Livermore, CA 94550

\end{center}

\vspace{0.3in}

\medskip
We consider the formation of horizon-size primordial black holes (PBH's)
from pre-existing density fluctuations during cosmic phase transitions.
It is pointed out that the formation of PBH's 
should be particularly efficient during the QCD
epoch due to a substantial reduction of pressure forces during
adiabatic collapse, or equivalently, a significant decrease in the
effective speed of sound during the color-confinement transition. 
Our considerations imply that
for generic
initial density perturbation spectra
PBH mass functions are expected to exhibit a pronounced peak on the 
QCD-horizon mass scale
$\sim 1 M_{\odot}$. This mass scale is roughly coincident with
the estimated masses for compact objects recently observed 
in our galactic halo by the MACHO
collaboration. 
Black holes formed during the QCD epoch may offer an attractive explanation
for the origin of halo dark matter evading possibly problematic
nucleosynthesis and luminosity bounds on baryonic halo dark matter.

\medskip
\centerline{PACS number(s): 97.60.LF, 98.35.+d, 12.38.AW, 98.35.Gi}

\newpage
\pagestyle{plain}

Ever since the early works by Zeldovich, Novikov\,\cite{Zel}, and
Hawking\,\cite{Hawk} 
it is clear that only moderate deviations from a perfectly 
homogeneous Friedman
universe can lead to copious production of
PBH's at early epochs.
Initially super-horizon
size overdense regions can collapse and convert into PBH's when they enter
into the particle horizon. 
Once a fluctuation passes into the particle horizon it's subsequent
evolution is essentially a competition between pressure forces and gravity.
When the equation of state is hard, $p=\rho /3$ where $p$ is pressure 
and $\rho$ is
energy density, fluctuations with overdensities exceeding a critical value 
$(\delta\rho /\rho )\ \simge\ \delta_c^{\rm RD}\approx 1/3$ are
anticipated to form black holes whereas fluctuations with overdensities 
less than this
critical value are expected to disperse due to pressure 
forces\,\cite{Carr3}.
Here overdensities are specified at fluctuation horizon crossing
in uniform Hubble constant gauge.
A typical mass for PBH's formed during a radiation dominated era
(e.g. $p=\rho /3$) is of the order of the horizon mass.
The horizon mass is given by

\begin{equation}
M_H(T)\approx 1M_{\odot}\biggl({T\over 100{\rm MeV}}\biggr)^{-2}\biggl
({g_{eff}\over
10.75}\biggr)^{-{1\over 2}}\ ,
\end{equation}
where $T$ is cosmic temperature and $g_{eff}$ are the effective relativistic
degrees of freedom contributing to the Hubble expansion. The only numerical,
albeit schematic,
simulation of PBH formation to date\,\cite{Nade}
indicates that PBH's form with masses somewhat smaller than the horizon mass. 
It has been emphasized that black hole production from pre-existing adiabatic
fluctuations can be very efficient when there is a period 
during the 
evolution of the
early universe for which the equation of state is soft 
($p\approx 0$)\,\cite{Carr3,Khlo}.
In the seventies the possible overabundant production of PBH's had also
been used to rule out a prolonged soft Hagedorn equation of state at
$T\sim 100$MeV\,\cite{Chap}. Nevertheless with the discovery of quarks
such speculative \lq\lq dust\rq\rq -like eras were 
believed to only occur at 
temperatures in excess
of the electroweak breaking scale ($T\ \simge\ 100$GeV).
 
There are also various schemes for the spontaneous, phase-transition dynamics
related, generation of density perturbations on sub-horizon scales 
during cosmological first order transitions and the
concomitant production of PBH's.
In the context of the QCD confinement transition
Crawford \& Schramm\,\cite{Craw} have argued that
the long-range color force could lead to the generation of sub-horizon
density fluctuations which may turn into planetary sized black 
holes
even though the details of this mechanism remain to be explored.
Hall \& Hsu\,\cite{Hall} 
proposed the formation of PBH's during a first order QCD transition
from imploding, supercooled quark-gluon plasma bubbles. Katalini\'{c} \& Olinto
point out that the possible tendency of extreme focusing of inwardly 
traveling
sound waves in a quark-gluon plasma bubble, reminiscent of the process of
sono-luminescence, can lead to density fluctuations at the center of the 
bubble
sufficiently large for the production of PBH's. 
Note that these schemes predict PBH masses which are far below
the QCD-horizon mass.
In the following we will show that the universe effectively has a soft equation
of state during a first-order QCD transition which, depending on cosmic
initial conditions, may lead to the abundant formation of PBH's on the
QCD horizon mass scale.

A first-order color-deconfinement QCD phase transition is characterized
by the coexistence of high-energy density quark-gluon phase (quarks and gluons
with strong mutual many-body interactions) with low-energy density hadron 
phase
(mostly color-singlet pions for a gas at vanishing chemical potential) at a 
coexistence
temperature of the order of $T_c\sim 100$MeV\,\cite{Witt,Appl,Full}.
Within the simplistic bag model
pressure $p$, energy density $\rho$, and entropy density $s$ of the
hadronic (h) and quark-gluon (qg) phases may be written as \cite{DeGra,Full}

\begin{equation}
p_h(T)={1\over 3}g_h{\pi^2\over 30}T^4\quad ;\quad p_{qg}(T)={1\over 3}g_{qg}
{\pi^2\over 30}T^4-B\ ;
\end{equation}
\begin{equation}
\rho_h(T)=g_h{\pi^2\over 30}T^4\quad ;\quad \rho_{qg}(T)=g_{qg}
{\pi^2\over 30}T^4+B\ ;
\end{equation}
\begin{equation}
s_h(T)={4\over 3}g_h{\pi^2\over 30}T^3\quad ;\quad s_{qg}(T)={4\over 3}g_{qg}
{\pi^2\over 30}T^3\ .
\end{equation}
Here $g_h\approx 17.25$ and $g_{qg}\approx 51.25$ are the statistical weights
of relativistic particles for the individual phases, including the contributions
of pions and leptons for the hadronic phase and quarks, gluons, and leptons
for the quark-gluon phase. The bag constant $B$ acts effectively as a vacuum
energy density for the quark-gluon phase and accounts for the strong mutual
interactions of quarks and gluons.
For a first-order transition at coexistence temperature $T_c$,
the conditions of 
thermodynamic equilibrium are
the equality of pressure
$p=-(\partial E/\partial V)_S$ and temperature 
$T=(\partial E/\partial S)_V$ between hadronic phase and 
quark-gluon phase.

One may consider a region of quark/hadron matter sufficiently
larged to include material in both phases. 
The pressure, average energy density $\langle\rho\rangle$,
and average entropy density $\langle s\rangle$ of such 
a region of 
quark/hadron matter are
\begin{equation}
p=p_{qg}(T_c)=p_{h}(T_c)\ ,
\end{equation}
\begin{equation}
\langle\rho\rangle = f_{qg}\rho_{qg}(T_c)+(1-f_{qg})\rho_{h}(T_c)\ ,
\end{equation} 
\begin{equation}
\langle s\rangle = f_{qg}s_{qg}(T_c)+(1-f_{qg})s_{h}(T_c)\ ,
\end{equation} 
where $f_{qg}$ is the
fraction of space permeated by quark-gluon phase. 
An adiabatic compression of quark/hadron matter in a state
of equilibrium phase coexistence induces a conversion of low-energy density
hadron phase into high-energy density quark-gluon phase
($f_{qg}$ rises). During this process the average
energy density increases while pressure and temperature remain
constant.
The pressure response of quark/hadron matter
to slow adiabatic expansion, compression, or collapse is 
therefore negligible. This may be expressed by 
defining an
effective speed of sound for quark/hadron matter in a state of phase mixture, 
such that in thermodynamic equilibrium

\begin{equation}
v_S^{eff}=\sqrt{\biggl({\partial p\over\partial
\langle\rho\rangle}\biggr)_S}= {0}\ ,  \label{eq:vs}
\end{equation}
holds exactly.
Eq.~\ref{eq:vs} implies that the universe is Jeans unstable during
the quark/hadron transition
for scales much smaller than the horizon length. In contrast, the Jeans
length during \lq\lq ordinary\rq\rq\ radiation dominated eras 
$(v_S=1/\sqrt{3})$
is of the order of the horizon length.

One may wonder if during a cosmic QCD phase transition thermodynamic
equilibrium, in particular, constant pressure and temperature are maintained.
Consider, for example, a region of quark/hadron matter at $T_c$. 
Upon compression
or collapse such a region could, in principle, superheat which would yield
a pressure response, such that $v_S^{eff}\neq 0$. Superheated quark/hadron
matter may cool by either the growth of existing quark-gluon phase or 
the nucleation
of new quark-gluon bubbles. 
One may estimate the amount of superheating $\eta=(T-T_c)/T_c$ at which 
heating due to adiabatic collapse
is balanced by cooling due to the nucleation 
of critically-sized quark-gluon bubbles

\begin{equation}
{\langle\rho\rangle\over t_H}\simeq 30 {\sigma^3T_c^4\over
L^2\eta^3}\exp{\biggl(-{16\pi\over 3}
{\sigma^3\over L^2\eta^2 T_c}\biggr)}\ ,  \label{eq:etan}
\end{equation}
where $t_H$, $L$, and $\sigma$ are Hubble time, latent heat
of the
transition, and surface free energy of the phase boundary, respectively, 
and we have assumed
$\eta\ll 1$. The right-hand-side of Eq.~\ref{eq:etan} gives the cooling rate
from nucleation of new phase with the exponent the change in free energy
due to the spontaneous appearence of a critically-sized bubble of
quark-gluon phase divided by the temperature\,\cite{Full,Igna},
whereas the left-hand-side of Eq.~\ref{eq:etan} is simply the rate of
change in energy density during collapse in the absence of cooling. 
This latter rate is approximately given by the
energy density of the mildly non-linear
fluctuation $\sim \langle\rho\rangle$ over the gravitational collapse time scale
$\sim t_H$.
Quite independent of the prefactors
cooling is efficient when the exponent in Eq.~\ref{eq:etan} is approximately
10-20, which yields $\eta\simeq \sigma^{3/2}/LT_c^{1/2}$. Typical parameters
of the phase transition have been estimated to be approximately 
$\sigma\simeq 0.02 - 0.1T_c^3$\,\cite{Alco1,Huang,Kaja,Iwa1}, 
$L\approx 2-15 T_c^4$\, \cite{Iwa2,DeTar,Blum,Bern} 
with $T_c\approx 100$MeV. This implies that
quark/hadron matter can not sustain superheating by more
than $\eta\approx 10^{-3}-10^{-5}$ during adiabatic collapse. 

These considerations illustrate that the pressure response of quark/hadron matter
to adiabatic compression, or equivalently the effective speed of sound, is
dependent on the amplitude and time scale of the compression process. 
For example,
small-amplitude, $(\delta T/T)\simle \eta$, sound waves in pure hadronic phase,
or quark-gluon phase, at $T_c$ will propagate with the ordinary
speed of sound $v_S=(\partial p/\partial\rho)_S^{1/2}\sim 1/\sqrt{3}$. 
However, the pressure
response  of \lq\lq mixed\rq\rq\ quark-gluon and hadron phase to significant
adiabatic collapse, $(\delta\langle\rho\rangle /\langle\rho\rangle )\sim 1$, will
be reduced substantially by a factor $\sim\eta$ 
when compared to the 
pressure
response of \lq\lq ordinary\rq\rq\ relativistic matter to collapse.
Note that the anamoly in the speed of sound during a first-order
QCD transition, which was independently discovered by
Schmid et al.\cite{Schmid}, may have interesting implications for the growth
of sub-horizon size initial density perturbations\cite{Schmid}.

The universe has effectively a soft 
equation of state during the QCD transition when considering the pressure
response to gravitational collapse of density fluctuations.
It is interesting to note that the evolution of average energy density 
$\langle\rho\rangle$ 
as a function of scale factor $R$ during a
first-order QCD transition is given by
\begin{equation}
\langle\rho\rangle (R)= \biggl({R_0\over R}\biggr)^3\bigl[\rho_{qg}+{1\over 3}\rho_h
\bigr]-{1\over 3}\rho_h\ ,
\end{equation}
which yields for the time evolution
of $R$ approximately $R(t)\sim t^{2/3}$, akin to a dust-like era.
Nevertheless, the duration of this dustlike phase is brief.
By using the bag model we may derive for the ratio of scale factors at 
the
beginning of the transition, $R_0$, to the value at 
completion of the
transition, $R_1$, from conservation of entropy
$(R_1/R_0)=({g_{qg}/g_{h}})^{1\over 3}\approx 1.44\ ,$
which implies that the duration of the transition
is of order of the Hubble time at the QCD epoch.
The duration of the transition may be even substantially shorter than
this estimate if the latent heat is smaller than the bag model
value, $L\approx 15T_c^4$, as indicated by most lattice QCD simulations
\cite{Iwa2,DeTar,Blum,Bern}.

Moderate amplitude, overdense fluctuations entering into the horizon
during the QCD transition will experience an almost complete reduction in
dispersing pressure forces for approximately the duration of the transition.
Such fluctuations will have a fraction of a Hubble time longer to collapse
unhindered by pressure forces
than their counterparts passing into the horizon during 
\lq\lq ordinary\rq\rq\ radiation
dominated eras. Therefore the demand on fluctuation
amplitude $\delta_c^{\rm QCD}$ for successful formation of a black
hole during the QCD transition should be lessened when compared
to the analogous quantity $\delta_c^{\rm RD}$ ,
\begin{equation}
\delta_c^{\rm QCD}<\delta_c^{\rm RD}\approx {1\over 3}\ .
\end{equation}
For almost scale-invariant initial adiabatic perturbations of the Harrison-Zeldovich
type, which have approximately equal amplitudes at horizon crossing on all scales, 
PBH's would form more abundantly during the QCD epoch than during radiation
dominated epochs
leading to a peak in the
PBH mass function at $M_{BH}\sim 1M_{\odot}$.
Note that a peak in PBH mass functions on a given mass scale 
may also result from non-trivial
initial
fluctuation spectra which, however, 
require fine-tuning of initial conditions\,\cite{Ivan, Yoko}.

For first order transitions which are much longer than the
QCD transition the computation of an analogous critical 
overdensity $\delta_c$ may be well
approximated by using a spherical top-hat model for the evolution
of the fluctuation\,\cite{Khlo1},
since there are no pressure forces between fluctuation and
environment.
For such transitions $\delta_c$ may be substantially
below $\delta_c^{\rm RD}$ for non-rotating fluctuations. However, in the
case of a first-order QCD transition we only expect mild (order unity)
decrease in $\delta_c^{\rm QCD}$ when compared to $\delta_c^{\rm RD}$ since the
epoch of phase coexistence is short. For successful PBH formation during the
QCD era fluctuation and environment may well exist in different regimes
during the pre-PBH formation evolution, for
example, the overdense fluctuation may be in 
a quark-gluon/hadron mixture in phase coexistence at $T_c$
whereas the environment may be purely in hadronic phase 
at $T\simle T_c$. In this case
the reduction in pressure gradients is only partial. Nevertheless, since pressure
gradients are always reduced a decrease in $\delta_c^{\rm QCD}$ compared to
$\delta_c^{\rm RD}$ seems inevitable.
A reliable estimate of $\delta_c^{\rm QCD}$ would have to be obtained with the help of a
numerical general-relativistic hydrodynamics code, using
an appropriate metric which asymtotically approaches the
Robertson-Walker metric in regions far away from a fluctuation,
and under the inclusion of detailed modeling of the QCD equation of state,
in regimes at, but also below and above, the transition point.

We have so far restricted our attention
to the QCD transition
and furthermore
implicitly assumed that color-confinement in the early universe
proceeds via a first order transition. Lattice gauge simulations are still not
conclusive as to the order of the QCD transition, mainly due to finite resolution
effects and the difficulties associated with simulating bare quark 
masses\,\cite{DeTar,Iwa3,Kana}.
We wish to stress that a possible reduction of the effective speed of sound
may be a generic feature of cosmic phase transitions and may not
necessarily be tied to the character of a transition.
In fact, a reduction in the speed of sound of order 10-20\% for a few Hubble times
does occur during the cosmic $e^+e^-$-annihilation\,\cite{Jed}
(the nature of the $e^+e^-$-annihilation is very different
from that of a first-order QCD color-confinement transition). This effect of pressure
reduction during phases of $e^+e^-$-creation and -annihilation is also well known
in stellar evolution calculations, commonly referred to as the
pair-instability, and relates to the conversion of relativistic energy density
(photons) to $e^+e^-$- rest mass energy density. We note that there is the
possibility of an enhancement in PBH formation on the $e^+e^-$-annihilation
horizon mass scale of approximately, $M\sim 10^5M_{\odot}$.

PBH's formed during the QCD epoch (or any other early epoch)
may contribute a significant
fraction $\Omega_{BH}$ to the closure density today if only a tiny fraction
$\epsilon$ of the radiation energy density at the QCD epoch
is converted into black hole mass density

\begin{equation}
\Omega_{BH}=5.8\times 10^7\epsilon (T)\biggl({T\over 100{\rm MeV}}
\biggr)\biggl({g_{eff}\over 10.75}\biggr)h^{-2}\ .
\end{equation}
Here $h$ is the Hubble constant in units of 100 km s$^{-1}$ Mpc$^{-1}$. 
This is because radiation energy density redshifts as $1/R^4$ whereas black hole mass
density redshifts as $1/R^3$ during the subsequent expansion of the universe.
Production of PBH's during the QCD epoch would also lead to the
spontaneous generation of isocurvature perturbations on super-horizon scales,
even though this isocurvature component is not expected to play a role in the
formation of large-scale structure unless $M_{BH}\,\simge\, 10^5M_{\odot}$
\,\cite{Carr4,Press}.
It is, however, interesting to note that PBH's form
on the peaks of the underlying adiabatic perturbations and PBH number density is
therefore strongly correlated with the adiabatic density fluctuations.

The efficiency $\epsilon (T_c)$ for PBH formation
during the QCD transition can be obtained from the statistics
of the initial density perturbations and from $\delta_c^{\rm QCD}$.
It is given by the fraction of QCD horizon volumes which are overdense
by more than $\delta_c^{\rm QCD}$,
\begin{equation}
\epsilon (T_c)=\int_{\delta_c^{\rm QCD}}^{\infty} f(\delta ,T_c)d\delta\
.\label{eq:101}
\end{equation}
Here $f(\delta ,T)$ is the probability distribution to find a 
horizon volume
overdense by $\delta =(\delta\rho /\rho)$, normalized such that
$\int_{-\infty}^{\infty}fd\delta =1$.
Currently favored mechanisms for the generation of primordial density
perturbations involve quantum fluctuations of scalar fields which drive an
extended inflationary period of expansion in the very early universe
($T\simge 1$TeV). Many of such models predict a Gaussian probability
distribution
\begin{equation}
f={1\over\sqrt{2\pi}}{1\over\sigma (M)}exp{\biggl(-{1\over 2}{\delta^2\over\sigma^2(M)}
\biggr)}\ . \label{eq:102}
\end{equation}
with approximate variance
\begin{equation}
\sigma (M)\approx 5\times 10^{-6}\biggl({M\over 5\times 10^{23}
M_{\odot} h^{-1}}\biggr)^{(1-n)\over 6}\ , \label{eq:103}
\end{equation}
where $n$ is a spectral index\cite{Carr5}. 
Simple inflationary scenarios predict equal
perturbation amplitudes for radiation and pressureless matter (CDM).
The reader be advised that the mass scale $M$ in Eq.~\ref{eq:103}
denotes the horizon mass in CDM only. For the QCD epoch, and assuming
a closed universe, this mass scale is $M_{\rm CDM}^H\approx 3\times 10^{-8}
M_{\odot}(T/100{\rm MeV})^{-3}h^2$. The simplest
inflationary models predict scale-invariance, $n=1$\,\cite{Kolb}. There are,
however, inflationary scenarios which predict the generation of blue
spectra\,\cite{Lids} ($n>1$). 
Using Eq.~\ref{eq:101}-Eq.~\ref{eq:103} and assuming
$\delta_c^{\rm QCD}$ in the range $\delta_c^{\rm QCD}=0.05 - 0.2$, 
one can infer that for spectral index in the
range $n=1.67-1.79$ cosmologically significant PBH formation during the
QCD epoch ($\Omega_{\rm PBH}\sim 1$) may occur.
This estimate asesses the sensitivity of the required spectral index for
significant QCD black hole formation on the undetermined $\delta_c^{\rm QCD}$.
We note that in Gaussian models a $\delta_c^{\rm QCD}$ only somewhat
smaller than $\delta_c^{\rm RD}$ results in PBH formation
essentially only on the QCD scale due to the steep decrease of $f$ with $\delta$.

A spectral index as large as $n=1.7$
is incompatible with observed cosmic microwave background
radiation (CMBR) anisotropies\,\cite{Benn} and spectral distortions\,\cite{Hu},
which imply $n\simle 1.5-1.6$. Moreover, the required spectral index for
significant QCD black hole formation may even exceed the above estimate
when non-Gaussian, skew-negative features, 
resulting for most inflationary scenarios
producing blue spectra, are taken into account\cite{Bull}.
This constraint may be circumvented in two ways. First, it may be that
the effective spectral index increases with decreasing mass scale. Cosmologically
significant PBH formation at the QCD scale may then still be compatible
with the CMBR limits, since those limits are derived on scales much larger
than the QCD horizon scale. However, in this case the primordial density 
perturbation
spectrum has to be such that constraints derived from PBH formation
on mass scales smaller than the QCD-horizon mass are not violated\,\cite{Carr1}.
These limits may be fairly stringent, in particular, when the reheating
temperature after an inflationary epoch is high $(T_{RH}\gg 1{\rm TeV})$ and/or
the inflationary epoch is followed by a prolonged reheating period.
As a second possibility it may be that the initial density perturbations are 
non-Gaussian\,\cite{Alle,Kibl} and conspire to have a skew-positive tail,
such that a fraction of approximately $\sim 10^{-8}$ horizon volumes are overdense
by more than $\delta_c^{\rm QCD}$. 

Our findings are particularly interesting in light of the 
recent observations of the MACHO collaboration\,\cite{Alco} that a
significant fraction of the dark matter in the galactic halo may be composed of
massive compact objects.
Even though the statistics of gravitational microlensing events is still poor
there is accumulating evidence for a sharp cutoff on the 
distribution of masses of compact 
halo objects, such that essentially all compact halo objects have
$M\,\simge\, 0.1M_{\odot}$ for a standard halo model.
The identification of these objects as red or white dwarfs may be problematic.
Searches for a dwarf population in our halo with the Wide Field Camera on the 
Hubble Space Telescope yield fairly stringent limits on the fraction of mass
contributed by dwarfs to the halo\,\cite{Flynn,Graff}. 
In the case of an abundant halo white dwarf population one may also have
to address the problem of overproducing metals.

In summary, we have argued that PBH formation from pre-existing adiabatic
fluctuations may be particularly efficient during cosmic phase transitions
and periods of particle annihilation
due to a softening of the equation of state. In the case of the QCD
color-confinement transition we have found that PBH's may form abundantly
on the mass scale $M_{BH}\sim 1M_{\odot}$, which is surprisingly close
to the inferred masses of compact objects recently discovered 
in our galactic halo by the MACHO
collaboration\,\cite{Alco}. A peak in the mass function of compact halo dark matter
may, in future, be observationally verified by gravitational microlensing
experiments. PBH's formed during the QCD epoch provide a natural explanation
for such a peak, and furthermore evade possibly problematic bounds on 
baryonic compact halo dark matter.

I am grateful to C. R. Alcock, G. M. Fuller, and A. Olinto for encouragement
and valuable discussions. This work was performed under the auspices of the
US Department of Energy by the Lawrence Livermore National Laboratory under contract
number W-7405-ENG-48 and DoE Nuclear Theory grant SF-ENG-48.

\end{document}